\documentclass[twocolumn,showkeys,preprintnumbers,amsmath,amssymb]{revtex4}
\usepackage{graphicx}% Include figure files
\usepackage{dcolumn}% Align table columns on decimal point
\usepackage{bm}% bold math
\usepackage{amsmath}
\usepackage{epstopdf}
\begin{document}
%\preprint{APS/123-QED}
\title{Wetting and Diffusion of Water on Pristine and Strained Phosphorene}% Force line breaks with \\
\author{Wei Zhang$^{1}$}
\author{Chao Ye$^{1}$}
\author{Lin Bi$^{2}$}
\author{Zaixing Yang$^{3}$}%
\author{Ruhong Zhou$^{2,3,4}$}%

\affiliation{%
$^1$Bio-X Lab, Department of Physics, and Soft Matter Research Center, Zhejiang University, Hangzhou 310027, China\\
$^2$Computational Biology Center, IBM Thomas J. Watson Research Center, Yorktown Heights, NY 10598, USA\\
$^3$Institute of Quantitative Biology and Medicine, SRMP and RAD-X, Collaborative Innovation Center of Radiation Medicine of Jiangsu Higher Education Institutions, Soochow University, Suzhou 215123, China\\
$^4$Department of Chemistry, Columbia University, NY 10027, USA}

\date{\today}% It is always \today, today,

\begin{abstract}
Phosphorene, a newly fabricated two-dimensional (2D) nanomaterial, have exhibited promising application prospect in biology. Nonetheless, the wetting and diffusive properties of bio-fluids on phosphorene are still elusive. In this study, using molecular dynamics (MD) simulations, we investigated the structural and dynamic properties of water on pristine and strained phosphorene. The MD simulations illustrated that the diffusion of water molecules on the phosphorene surface is anisotropic, while strain-enhanced diffusion is clearly present which arises from strain-induced smooth of the energy landscape. The contact angle of water droplet on phosphorene exhibited a nonmonotonic variation with the transverse strain. The structure of water on transverse stretched phosphorene was demonstrated to be different from that on longitudinal stretched phosphorene. Moreover, we discovered that the contact angle of water on strained phosphorene is proportional to the quotient of longitudinal and transverse diffusion coefficients of interfacial water. These findings would offer helpful insights in potential ways of manipulating the wetting and transport of water at nanoscale, and in future bio-applications of phosphorene.
\end{abstract}

%\pacs{82.70.Dd, 68.05.Gh, 83.85.Jn, 05.40.-a.}% PACS, the Physics and Astronomy
%Use showkeys class option if keyword
\maketitle

\section{\label{sec:level1}Introduction}
As a newly fabricated 2D nanomaterial composed of phosphorus atoms, phosphorene possesses a direct band gap which makes it a natural semiconducting nanomaterial \cite{LiuW14,LiuH15}. The specific properties of phosphorene has made its potential applications in biology possible \cite{LingX15,ZhangW15,SunZ15}. Notably, using a few layers of phosphorene, researchers have successfully fabricated the field-effect transistor \cite{LiL14,LiuH14}, which is a desirable device for biodetection and biosensing \cite{ImH07,Martinez09,TianB10}. Compared to graphene, phosphorene is more biological friendly due to its less disruption to proteins \cite{ZhangW15}. Moreover, as photothermal agents, black phosphorus (multiple layers of phosphorene) quantum dots have exhibited excellent performances in killing C6 and MCF7 cancer cells \cite{SunZ15}. Extensive bio-applications of this nanomaterial need more insights in the wetting and diffusive properties of bio-fluids in contact with the phosphorene.

The structure and dynamics of interfacial water at various nanomaterial surfaces are of fundamental importance for developing nanomaterials¡¯ potential applications in biology and nanofluidics. The structure of water is perturbed heavily near surfaces which have the potential to affect water diffusion \cite{Kim13,LiQ15} and proteins¡¯ adsorption \cite{Peter14}. During recent years, the wetting and diffusive properties of water in contact with two-dimensional (2D) nanomaterials, such as graphene \cite{LiH12,Taherian13,LiX13,WeiN14,ChenZ13}, boron-nitride sheets \cite{LiH12,Gordillo11,Tocci14}, WS2 and MoS2 \cite{Chow15,Kozbial15} etc., have been widely examined. These studies have strongly promoted applications of these 2D nanomaterials in biology and nanofluidics. The structure and dynamic of water on phosphorene, a new member of 2D nanomaterials, are still elusive.

Another promising aspect of phosphorene is its great mechanical flexibility \cite{WeiQ14}, due to the hexagonally arranged phosphorus atoms and the subsequently formed puckered honeycomb structure inside the monolayer of phosphorene. A computational result based on first-principles showed that single-layer phosphorene could actualize tensile strain up to $\sim$0.5 \cite{Jiang14}. It has also been demonstrated that the electrical \cite{FeiR14,HuT14,GeY15,LiY14}, optical \cite{Cakir14,Mehboudi15}, thermoelectric \cite{QinG14,LvHY14,Konabe15} and mechanical \cite{Jiang14,HuT14,Elahi15} properties of phosphorene could be modified upon mechanical strain. The puckered structure of phosphorene brings anisotropy and negative Poisson ratio \cite{Jiang14,QinG14}. The strain also gives rise to phosphorene's transition between metal and semiconductors \cite{LiY14,QinG14,Elahi15}. Therefore, what kind of changes in the structure and dynamics of water on strained phosphorene would occur are naturally topics of interest.

In this study, using molecular dynamics (MD) simulations, we measured the contact angle, diffusion coefficient and distribution of water on pristine and strained phosphorene. We found that the diffusion of water molecules at the surface of phosphorene is anisotropic, and strain-enhanced water diffusion is clearly present. The structure and wetting of water on transverse stretched phosphorene differs from that on longitudinal stretched one. The contact angle of water droplet nonmonotonously changes with the transverse strain, but it nearly linearly changes with the longitudinal strain. Also, the distribution of water near the surface of phosphorene exhibited obvious changes when the transverse strain increased, while the distribution of water barely changed when the longitudinal strain was imposed on phosphorene. Additionally, we found that the contact angle of water on strained phosphorene is proportional to the quotient of longitudinal and transverse diffusion coefficients of interfacial water. The dispersion energy and free energy profile of water were calculated to interpret the above phenomena.

\section{\label{sec:level1}Methods}
MD simulations were performed to generate strained phosphorene, as well as to explore the wetting and diffusive properties of water on pristine and strained phosphorene. The Gromacs package 4.5.7 \cite{Pronk13} and OPLS-AA force field \cite{Jorgensen96} were used for the simulations. The SPC/E model was used to model water molecules. Phosphorene with a dimension of 156 {\AA} $\times$ 173 {\AA} was chosen for this study. The phosphorus atoms were modeled as uncharged Lennard-Jones particles. The depth of potential well $\varepsilon_{pp}$, cross sections $\sigma_{pp}$ and bond strength constants are set at 0.400 kcal mol$^{-1}$, 3.33 {\AA} and 297 kcal mol$^{-1}$ {\AA}$^{-2}$ \cite{Ballone04}, respectively.

\begin{figure*}\label{Fig1}
\includegraphics[width=0.7\textwidth]{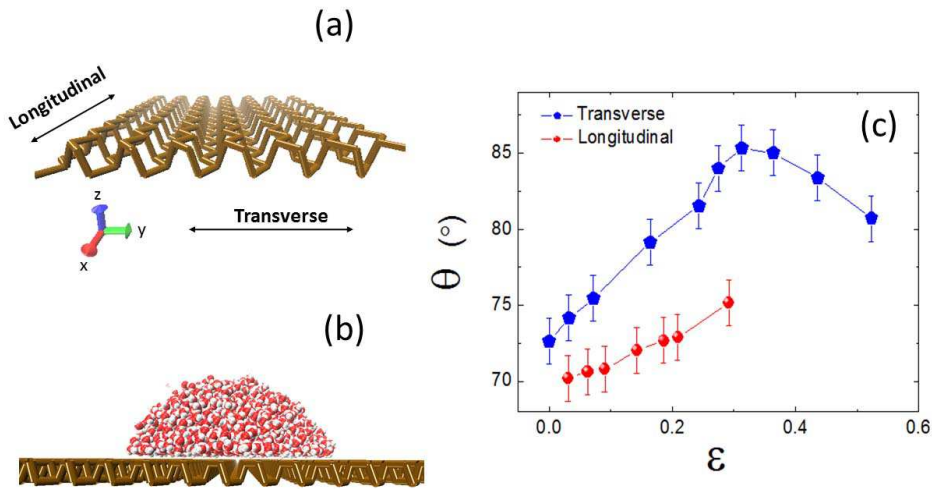}% Here is how to import EPS art
\caption{(a) Perspective view of transverse and longitudinal directions upon which strains were imposed. (b) A snapshot of water droplet on phosphorene ($\varepsilon = 0$) in the end of the MD simulation. (c) The water contact angle $\theta$ as a function of the transverse and longitudinal strain.}
\end{figure*}

Using the pulling code of Gromacs, we produced phosphorene under various strain conditions. Note that the strain being discussed in this study only refers to tensile strain. Phosphorene was therefore stretched along two typical direction: transverse (perpendicular to the pucker) and longitudinal (parallel to the pucker), as shown in Fig.1(a). Phosphorene with strain $\varepsilon = 0$ corresponds to the pristine phosphorene. Figure S1 in the Supporting Information shows four representative configurations of strained phosphorene. The initial water droplet which was configured as a cubic and consists of 2828 molecules was set onto phosphorene under various strain condition. During the first few nanoseconds of the simulation, the water droplet gradually converted from a cubic into a hemisphere, as illustrated in Fig. 1(b). We chose 10 morphologies of phosphorene with different degrees of transverse strain, as well as other 10 phosphorene morphologies with different degrees of longitudinal strain for the interaction with water droplet. Each phosphorene-water system was simulated for 16 ns, and the trajectory of the last 1 ns was extracted for further analysis. The strained phosphorene was fixed throughout the simulation. The particle-mesh Ewald method \cite{Darden93} with a grid spacing of 1.2 {\AA} was applied to simulate the long-range electrostatic interactions, and a typical cutoff 10 {\AA} was applied for the van der Waals interactions. All simulations were performed in an NVT ensemble at a constant temperature of 298 K by using v-rescale thermostat \cite{Bussi07}.

The water contact angle $\theta$ was measured by fitting the time-averaged liquid-vapor interface \cite{Werder03}. The liquid-vapor interface is defined as the contour with half of the bulk density, while the number density of water droplet was calculated by the time-averaged spatial mesh with a grid spacing of 0.5 {\AA}. The liquid-vapor interface was then fitted into an arc, while ¦È was calculated as the angle of contingence at the liquid-solid interface (refer to Figure S2 in the Supporting Information).

\section{\label{sec:level1}Results}

The Young's modulus of phosphorene is sensitive to directions upon which strains were imposed, due to the nanomaterial's anisotropic nature. Figure S3 in Supporting Information illustrates the relationship between the strain at transverse and longitudinal directions and the pull force exerted along these directions. The slope of transverse strain as a function of pull force is clearly steeper than that of longitudinal strain, indicating less stiffness in phosphorene when stretched in the transverse direction. These results consist with previous calculations based upon first-principles density functional theory \cite{WeiQ14,Jiang14}.

The strains imposed upon the nanomaterial affect not only phosphorene's electronic and mechanical properties, but the contact angle of water droplet on phosphorene as well. As shown in Fig. 1(c), the strains exerted in different directions have a distinct effect on the contact angle $\theta$. An increase of the transverse strain $\varepsilon_T$ causes $\theta$ to first increase, and then decrease after the peak is reached. At $\varepsilon_T$ $\sim$ 0.3, we observed the maximal contact angle $\theta_{max}$. However, for strains imposed upon the longitudinal direction, the relationship between $\theta$ and longitudinal strain rate $\varepsilon_L$ is clearly monotonic. A comparison between contact angles of water droplet on phosphorene under different directional strains also showed an overall larger contact angles under transverse than longitudinal strains.

Along with the differences in water contact angles, the transverse and longitudinal strains also cause different deformations of the phosphorene surface. While strain in the transverse direction effectively flattens the pucker of phosphorene, longitudinal strain, however, has little effect on the bending structure of the pucker ring. The distinct deformation of the monolayer of phosphorene leads to different interactions between water droplet and phosphorene. Figure S4 in Supporting Information illustrates the interaction energy $E_{WS}$ between water droplet and the substrate phosphorene as a function of the strain along transverse and longitudinal directions. The energy curves exhibit similarity to that of the contact angle as shown in Fig. 1(c). Due to this similarity, we examined the relationship between $\theta$ and $E_{WS}$ by fitting the former with the latter variable. As shown in Fig. S5, there is a clear linear relationship between the two variables, thus inferring a major influence upon the water contact angle by the interactions between water droplets and the substrate.

\begin{figure}\label{Fig2}
\includegraphics[width=0.5\textwidth]{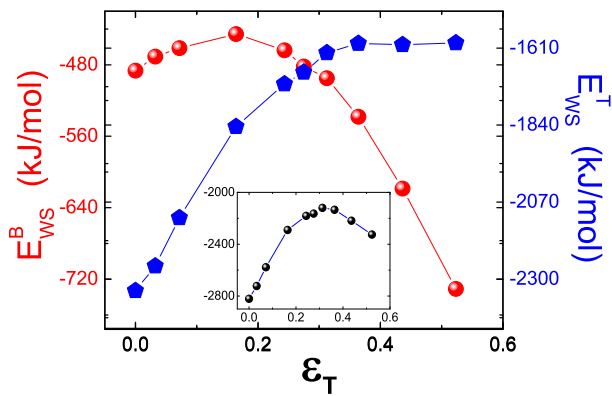}% Here is how to import EPS art
\caption{The interaction energy between water droplet and phosphorous atoms in bottom ($E^{B}_{WS}$ red circles) and upper ($E^{T}_{WS}$ blue pentagons) surface of the puckered surface of phosphorene as a function of the strain along transverse direction. The inset shows the combined interaction energy $E_{WS}=E^{B}_{WS}+E^{T}_{WS}$.}
\end{figure}

\begin{figure*}\label{Fig3}
\includegraphics[width=0.8\textwidth]{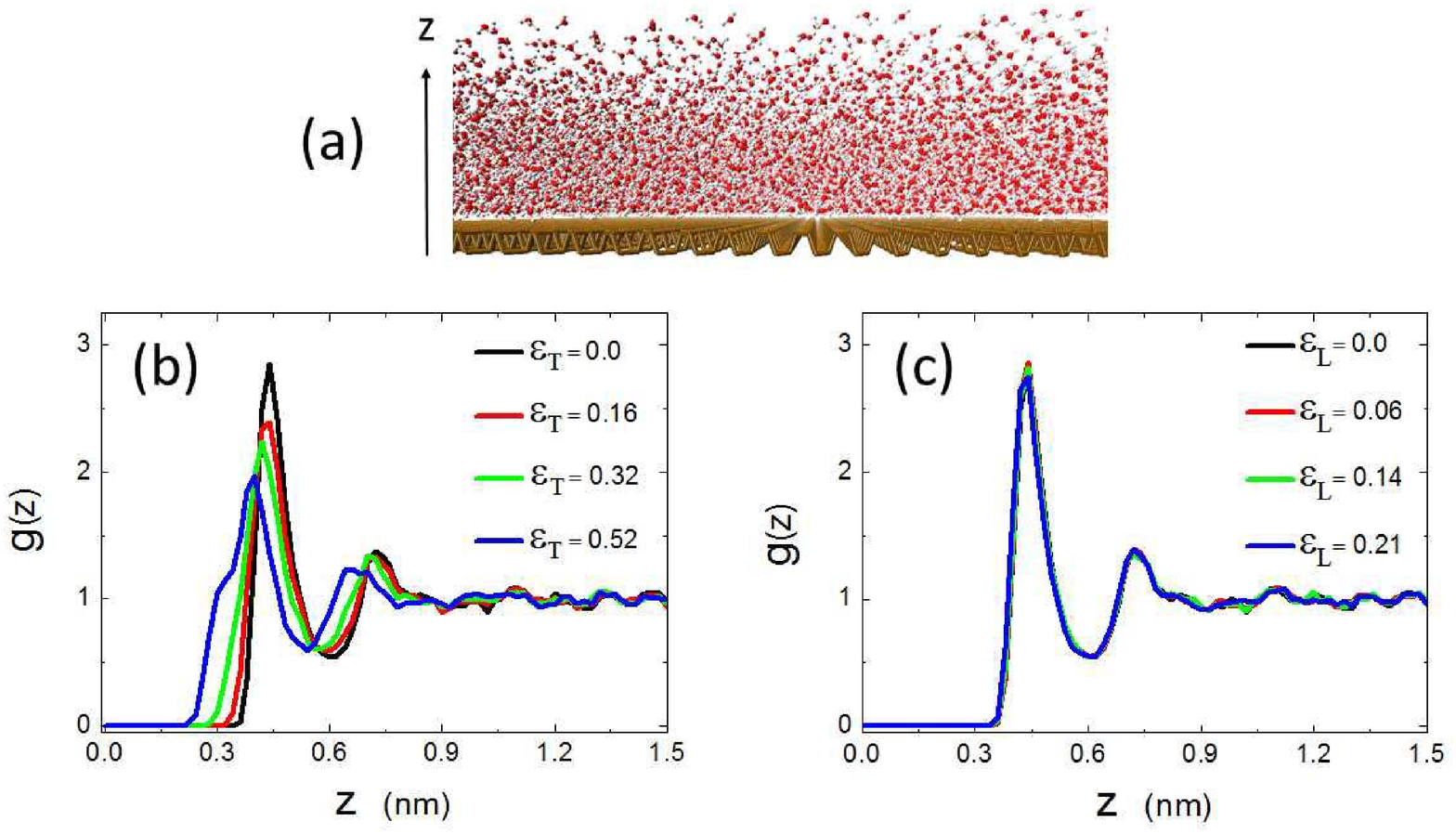}% Here is how to import EPS art
\caption{(a) A snapshot of the simulation system for studying the density distribution and diffusion of water molecules. (b) and (c) The density distribution function of oxygen atoms $g(z)$ as a function of the distance $z$ under different magnitudes of strains.}
\end{figure*}

The contact angle $\theta$ nonmonotonically changes with the transverse strain, which is primarily caused by nonmonotonic variations of the interaction energy $E_{WS}$. To analyze the influences of the transverse strain on $E_{WS}$ in detail, we calculate the interaction energy between water droplet and phosphorous atoms in bottom $E^{B}_{WS}$ and upper $E^{T}_{WS}$ surfaces of the puckered monolayer of phosphorene, respectively (see Fig. 2). Here $E_{WS}=E^{B}_{WS}+E^{T}_{WS}$. With the increase of the transverse strain $\varepsilon_T$, the density of atoms in upper surface decreases, thus $E^{T}_{WS}$ weakens remarkably. Meanwhile, the atoms in bottom surface approach the water droplet and its density also decreases. When $\varepsilon_T$ $< \sim$ 0.16, the decrease of the density of atoms in bottom surface dominates the interaction, and $E^{B}_{WS}$ weakens slightly. While $\varepsilon_T$ $>\sim$ 0.16, however, the approach of atoms in bottom surface dominates the interaction, and $E^{B}_{WS}$ enhances dramatically. The combined interaction effect of upper and bottom surfaces atoms with water droplet is that as increasing $\varepsilon_T$ the interaction energy $E_{WS}$ increases firstly and then decreases after $\varepsilon_T$ reaches $\sim$ 0.3, as shown in the inset of Fig. 2.

In addition to water contact angle, the strain may also affect the structure of water near phosphorene. In order to understand the structure of interfacial water molecules, we measured the density distribution function (DDF) of oxygen atoms along the direction normal to the surface (z axis). DDF $g(z)$ is defined as in Equation (1):

\begin{equation}
g(z)=\rho(z)/\overline{\rho},
\label{Eq1}
\end{equation}
where $\rho(z)$ is the density of oxygen atoms within a thin slice of height z parallel to the surface (the thickness of the slice is set at 0.2 {\AA}), and $\overline{\rho}$ is the mean density of oxygen atoms in the bulk area. Note that the zero point of z corresponds to the geometric center of phosphorene in z axis direction. An equilibrium system consisting of 25662 water molecules, as shown in Fig. 3(a), was used to study the density distribution and diffusion of water molecules. Figure 3(b) and (c) exhibited $g(z)$ of oxygen atoms near phosphorene under various transverse and longitudinal strains. Due to the dispersive interaction between phosphorous atoms and water molecules, the structure of water near phosphorene is considerably affected and shows double peak character. The double peaks of $g(z)$ in both Fig. 3(b) and (c) indicate the two-layer structure of water in the vicinity of phosphorene, with the first peak corresponding to the first layer of water. The density of oxygen atoms in the first water layer can be 2.8 times larger than that in the bulk. While $z$ $>\sim$ 1.0 nm, the density fluctuation disappears and the bulk density is recovered. Meanwhile, the difference between Fig. 3(b) and (c) is worth noting. In Fig. 3(b), as transverse strain increases, both the critical position $z_c$ where oxygen atoms appear, as well as the maximal DDF $g_{max}$ decreases, but the half-width of the first peak $W$ increases (see Table 1). On the other hand, the longitudinal strain showed almost no effect on $g(z)$, as shown in Fig. 3(c).

\begin{table}
\caption{\label{table1} The critical position $z_c$, the height $g_{max}$ and half-width $W$ of the first peak of the density distribution function (DDF) of oxygen atoms corresponding to various values of the transverse strain $\varepsilon_T$.}
\begin{ruledtabular}
\begin{tabular}{cccccccc}
$\epsilon_T$ & $z_c$ (nm) &  $g_{max}$  &  $W$ (nm) \\
\hline
0 & 0.36 & 2.85 & 0.08  \\
0.03 & 0.35 & 2.68 & 0.10  \\
0.07 & 0.34 & 2.64 & 0.10  \\
0.16 & 0.32 & 2.38 & 0.10  \\
0.24 & 0.30 & 2.26 & 0.12  \\
0.28 & 0.30 & 2.21 & 0.12  \\
0.31 & 0.28 & 2.24 & 0.12  \\
0.36 & 0.25 & 2.14 & 0.14  \\
0.43 & 0.24 & 2.08 & 0.15  \\
0.52 & 0.22 & 1.97 & 0.18  \\
\end{tabular}
\end{ruledtabular}
\end{table}

\begin{figure*}\label{Fig4}
\includegraphics[width=0.8\textwidth]{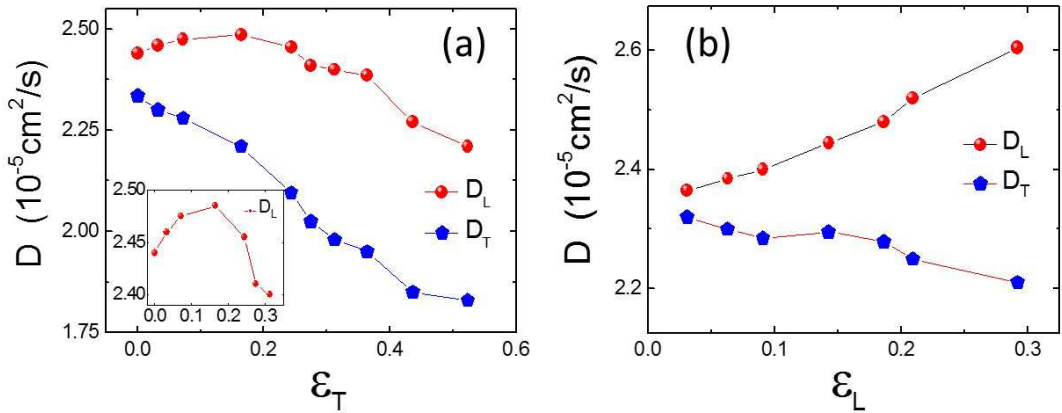}% Here is how to import EPS art
\caption{The diffusion coefficient $D$ of water molecules in the first layer as a function of (a) the transverse and (b) longitudinal strain.}
\end{figure*}

The above results demonstrated that the effects of transverse and longitudinal strain on the static structure of interfacial water are distinct, which can be partly attributed to the anisotropic mechanical properties of phosphorene. In the following discussion, we present the effects of the strain on the dynamic properties of interfacial water.

In order to examine the dynamic properties of liquid on phosphorene, we begin with the self-diffusion of water molecules, since it is considered to be the most representative dynamic property of liquid upon nanomaterials. The transverse and longitudinal self-diffusion coefficients $D_T$ and $D_L$ are derived from the following equations:

\begin{equation}\label{Eq2}
\begin{aligned}
D_T=\frac{1}{2N\tau}\sum_{i}\langle |y_i(t+\tau)-y_i(t)|^2 \rangle_{t},\\
D_L=\frac{1}{2N\tau}\sum_{i}\langle |x_i(t+\tau)-x_i(t)|^2 \rangle_{t},
\end{aligned}
\end{equation}
where $x_i (t)$ and $y_i (t)$ are the coordinates of the $ith$ water molecule at time $t$; $\tau$ is the lag time and N is the number of molecules included in this calculation. The angle brackets $\langle ... \rangle_t$ indicate an averaged calculation over a duration of time $t$. Figure 4 shows the transverse and longitudinal diffusion coefficient of water molecules near phosphorene with various strain conditions. It is clear from the separation of curves as shown in Fig. 4, that the diffusion of water molecules on pristine phosphorene is anisotropic, mainly due to the puckering surface morphology of phosphorene. Compared to motions in the transverse direction, a higher $D_L$ coefficient indicates that it is much easier for water molecules to move along longitudinal direction. The transverse diffusion coefficient $D_T$ decreases monotonically with increased transverse strain $\varepsilon_T$. The longitudinal diffusion coefficient $D_L$ increases while $\varepsilon_T$ is smaller than 0.16, and decreases after $\varepsilon_T$ rises above 0.16, as shown in Fig. 4(a) and its inset. The effects of longitudinal strain on the diffusion of interfacial water differ from that of transverse strain. As increasing longitudinal strain $\varepsilon_L$, $D_T$ decreases but $D_L$ increases. As shown in Fig. 4(b), $D_L$ increases from 2.37 to 2.61 ($10^{-5}$$cm^2/s$) in the measured range of $\varepsilon_L$. In contrast, the change of $D_T$ is not as significant. Clearly, the effects of $\varepsilon_T$ and $\varepsilon_L$ on the diffusion of interfacial water are all of an anisotropic manner.

\begin{figure*}\label{Fig5}
\includegraphics[width=0.9\textwidth]{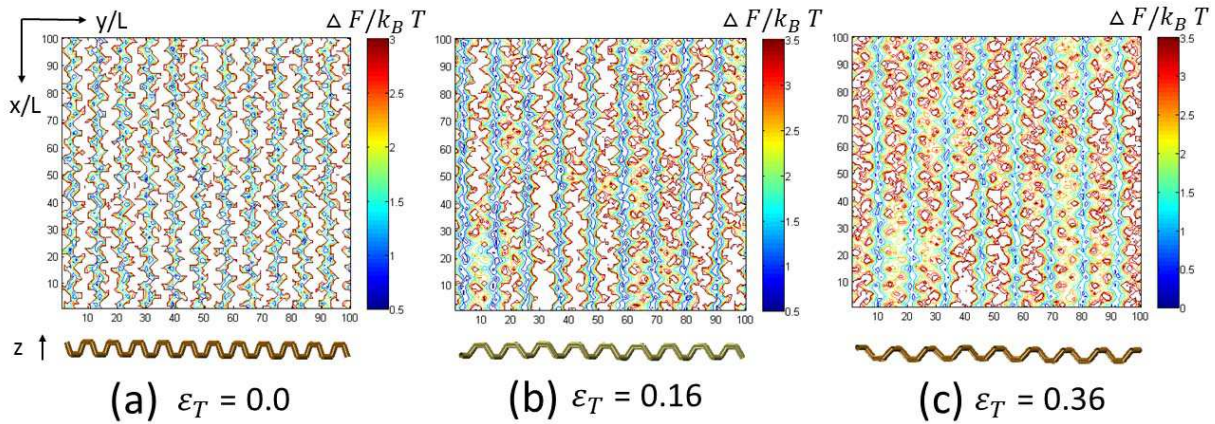}% Here is how to import EPS art
\caption{Free energy profile of water within the first layer $\Delta F(x,y)$ scaled by $k_BT$ for three values of $\varepsilon_T$. Here, the scaling parameter $L=0.05$ nm.}
\end{figure*}

To rationalize the diffusion behaviors of the interfacial water, we computed the free energy profile of water within the first layer $\Delta F(x,y)$ with the equation:

\begin{equation}\label{Eq3}
\Delta F(x,y)=-k_BTlnP_o(x,y).
\end{equation}
Here, $P_o (x,y)$ is the spatial probability distribution function of the oxygen atoms of water within the first layer at coordinate $(x,y)$. This approach has been previously applied to examine the friction of water on graphene and boron nitride \cite{Tocci14}. In Fig. 5, we present $\Delta F(x,y)$ as scaled by $k_B T$ under different transverse strains. The free energy profile $\Delta F(x,y)$ obviously exhibits a grooved trend of phosphorene surface, which is an indication of anisotropy. The coordinate with the minimal value of free energy appears in the region of the groove of phosphorene.

For pristine phosphorene ($\varepsilon_T=0$ and $\varepsilon_L=0$), the free energy profile $\Delta F(x,y)$ shows clear "zigzag" pattern in the grooved region (see Fig. 5(a)), which is very similar to famous "swallow gird". The maximal energy barrier for water molecules to translate along longitudinal direction is about 2 $k_B T$, while water molecules need to cross energy barrier of 3 $k_B T$ to move in transverse direction. It is more difficulty for interfacial water molecules to move in transverse direction compared to longitudinal translation. Thus, the diffusion coefficient of interfacial water along the longitudinal direction is larger than that along the transverse direction.

The strain $\varepsilon_T$ acts to broaden the interval of free energy ribbons (low-energy region along the groove) which correspond to groove width of phosphorene. The broadening of free energy ribbon interval constrains the diffusion of water molecules along transverse direction, since it introduces more difficulty in water molecules' crossing of broadened energy barrier. Meanwhile, the strain $\varepsilon_T$ increases the mean energy barrier for water molecules to move in transverse direction, thus hindering the motion of water molecules along the transverse direction. Consequently, the transverse diffusion coefficient $D_T$ decreases as $\varepsilon_T$ increases, as shown in Fig. 4(a).

With an increased strain $\varepsilon_T$, the "zigzag" form gradually disappears and the free energy ribbons become smooth (the low-energy region links in line), which is propitious to longitudinal diffusion of water molecules. Thus, the longitudinal diffusion coefficient $D_L$ increases when $\varepsilon_T$  is smaller than 0.16, as shown in the inset of Fig. 4(a). While $\varepsilon_T$ is larger than 0.16, the effects of phosphorus atoms at the bottom of the groove begin to dominate water's diffusion. As $\varepsilon_T$ increases $D_L$ decreases, due to the increased attraction of phosphorus atoms at the bottom of the groove (see Fig. 2).

As for the situations while phosphorene is under longitudinal stretching, the variations of the free energy profile $\Delta F(x,y)$ is different from those while phosphorene is under transverse stretching (see Fig. S6 in Supporting Information). The longitudinal strain $\varepsilon_L$ obviously attenuates "zigzag" pattern of free energy ribbon, which accounts for the enhancement of $D_L$ as $\varepsilon_L$ increases. However, unlike the effect of $\varepsilon_T$, the longitudinal strain $\varepsilon_L$ narrows the groove width, while shows little effect on its depth, which increases mean height of energy barrier but decreases its width. The increased energy barrier constrains the diffusion of water molecules along transverse direction. On the other hand, the decrease of the width of energy barrier is propitious to increase $D_T$. Under the combined influences of these two factors (the increased height and decreased width of energy barrier), as $\varepsilon_L$ increases, the transverse diffusion coefficient $D_T$ decreases slightly.

\section{\label{sec:level1}Discussion}

Though the longitudinal strain $\varepsilon_L$ obviously smoothes the free energy ribbon and increases the longitudinal diffusion coefficient of interfacial water, it has all most no effects on DDF g(z) (see Fig. 3(c)). The DDF g(z) in Fig. 3(b), however, obviously changes with the transverse strain $\varepsilon_T$, which is caused by the flattening of the puckered surface of phosphorene. The flattening of the surface of phosphorene leads to the decrease of the critical position $z_c$ and the increase of the space interval of free energy ribbon. The increase of the half-width $W$ may derive from the increased ribbon interval. The decrease of the height $g_{max}$ mainly arise from the weakened interaction energy between water and phosphorene. Consequently, DDF $g(z)$ is related not only to the dispersion energy between water and nanomaterials, but also to the distribution of water molecules on nanomaterials' surface.

\begin{figure}\label{Fig6}
\includegraphics[width=0.5\textwidth]{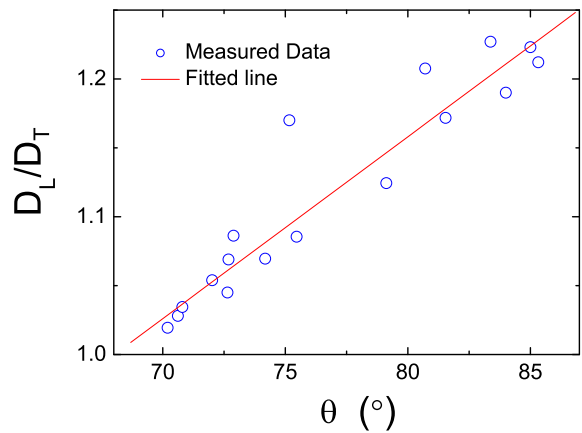}% Here is how to import EPS art
\caption{The quotient of $D_L/D_T$ as a function of the contact angle $\theta$ of water droplets. The fitting line has the form: $D_L/D_T=0.1+\theta\times0.013(^o)^{-1}$.}
\end{figure}

The quotient $D_L/D_T$, instead of $D_L$ or $D_T$ per se, exhibits a linear relationship with the contact angle $\theta$ of water droplet, as shown in Fig. 6. The variation of the width and depth of the groove of phosphorene caused by the strain directly affects the interaction energy $E_{WS}$ and the free energy profile $\Delta F(x,y)$. The contact angle $\theta$ and the diffusion coefficient of water molecules in the first layer are mainly determined by the interaction energy $E_{WS}$ and free energy profile $\Delta F(x,y)$, respectively. The free energy profile $\Delta F(x,y)$ jointly affects the transverse and longitudinal diffusion of water molecules. In this sense, the contact angle $\theta$ should be related to the community of the diffusion of water molecules along transverse and longitudinal direction. However at present, we have not yet found a quantitative interpretation regarding the relationship between the ratio $D_L/D_T$ and water contact angle $\theta$, which needs further investigation.

Changing the form of puckering surface of phosphorene by the strain could effectively enhance or attenuate the diffusion of interfacial water molecules, which might shed light on controlling/designing the motion of interfacial molecules. By controlling the strain, one could construct continuous diffusion (or wetting) gradient, which is of interest in artificial microscopic walk \cite{WangT15,Steimel14}. The anisotropic diffusion of water molecules near phosphorene may also affect phosphorene's motion in complex biological systems, which is of importance for phosphorene's potential bio-applications, such as localized bioprobes and drug delivery.

\section{\label{sec:level1}Conclusion}

In this study, we investigated the wetting and diffusive properties of water near both pristine and strained phosphorene with MD simulations. It was found that the pristine phosphorene is of weakly hydrophobicity, with a water contact angle of $\sim72^o$. As for the interactions between water droplet and phosphorene under different strain conditions, we discovered that the contact angle $\theta$ of water droplets firstly increases, then decreases as the transverse strain reaches the critical threshold of 0.3. However for longitudinal strain, the contact angle $\theta$ only increases monotonically as the strain increases. The changes in contact angles are mainly determined by the interaction energy between the water droplet and phosphorene. The structure of the interfacial water dramatically changes with the transverse strain $\varepsilon_T$, but the longitudinal strain $\varepsilon_L$ has almost no effect on water's structure.

While the diffusion of water molecules near pristine phosphorene surface is anisotropic, the longitudinal diffusion coefficient $D_L$  is larger than that ($D_T$) in the transverse direction. As the transverse strain $\varepsilon_T$ increases, $D_T$ decreases monotonically, while $D_L$ exhibits an inverted U-shaped curve. The longitudinal strain $\varepsilon_L$, on the other hand, causes $D_L$ to increase, and $D_T$ to decrease monotonically. We also calculated the free energy profile $\Delta F(x,y)$ so as to determine the main cause of variations in the diffusion of water molecules near phosphorene. It was shown that the smoothing of the energy landscape enhances $D_L$, and the increased energy barrier makes $D_T$ decrease. Last but not least, we found that the quotient $D_L/D_T$ is positively correlated with the contact angle $\theta$.

As a novel 2D nanomaterial, phosphorene has the potential to be applied extensively in biological systems in the future. It is hence necessary to investigate the wetting and diffusive properties of water on both pristine and strained phosphorene, so as to shed light on possible applications. Our study would potentially help with understanding and manipulating the wetting and diffusive properties of liquid on phosphorene, which is critical for phosphorene's application in the fields of biology and nanofluidics.

Acknowledgments:
This research is supported in part by the National Natural Science Foundation of China (Grant Nos. 11574268, 11504032, and 11474054), and China Postdoctoral Science Foundation (Grant Nos. 2015T80610 and 2014M560473).

%\begin{references}

%\newpage

%\end{references}

\end{document}